\documentclass[a4paper,12pt]{article}
\usepackage{cite}
\usepackage{amssymb}
\usepackage{multicol}
\usepackage{amsmath}
\usepackage{pgfplots}
\usepackage{mathrsfs} 
\usepackage{xfp} 
\usepackage[outline]{contour} 

\usepackage{epsfig}
\usepackage{graphicx}
\usepackage{slashed}
\usepackage{xcolor,color}
\usepackage{ulem}
\usepackage{subfigure}
\usepackage[toc,page]{appendix}
\usepackage[makeroom]{cancel}
\usepackage{tikz}
\usetikzlibrary{shapes.geometric,positioning,decorations.pathreplacing,decorations.pathmorphing,patterns,decorations.markings}

\usepackage{scalerel}
\usepackage[usestackEOL]{stackengine}

\def\dashint{\,\ThisStyle{\ensurestackMath{%
            \stackinset{c}{.2\LMpt}{c}{.5\LMpt}{\SavedStyle-}{\SavedStyle\phantom{\int}}}%
        \setbox0=\hbox{$\SavedStyle\int\,$}\kern-\wd0}\int}

\makeatletter
\renewcommand{\theequation}{\thesection.\arabic{equation}}
\@addtoreset{equation}{section}
\makeatother
\def\be{\begin{equation}}
\def\ee{\end{equation}}
\def\bea{\begin{eqnarray}}
\def\eea{\end{eqnarray}}
\def\bgn{\begin{align}}
\def\egn{\end{align}}
\def\nn{\nonumber \\}
\def\({\left(}
\def\){\right)}
\def\<{\left<}
\def\>{\right>}

\def\({\left(}
\def\){\right)}
\def\<{\left<}
\def\>{\right>}
\def\!{\right|}
\def\|{\left|}

\def\[{\left[}
\def\]{\right]}

\def\+{\bar}

\def\W{{\cal{W}}}

\usepackage{hyperref}
\hypersetup{colorlinks,%
  citecolor=blue,%
  linkcolor=blue,%
  pdftex}

\begin{document}

\begin{titlepage}
\vskip1cm
\begin{flushright}
\end{flushright}
\vskip0.25cm
\centerline{
\bf \large 
Discrete Bulk Spectrum in Jackiw-Teitelboim Theory
} 
\vskip0.8cm \centerline{ \textsc{
 Dongsu Bak,$^{ \negthinspace  a}$  Chanju Kim,$^{ \negthinspace b}$ Sang-Heon Yi,$^{\negthinspace a}$} }
\vspace{0.8cm} 
\centerline{\sl  a) Physics Department \& Natural Science Research Institute}
\centerline{\sl University of Seoul, Seoul 02504 \rm KOREA}
 \vskip0.2cm
 \centerline{\sl b) Department of Physics, Ewha Womans University,
  Seoul 03760 \rm KOREA}
\vskip0.4cm

 \centerline{
\tt{(\small dsbak@uos.ac.kr,\,cjkim@ewha.ac.kr,\,shyi704@uos.ac.kr})
} 
  \vspace{1.5cm}
\centerline{ABSTRACT} \vspace{0.65cm} 
{
\noindent
We argue that a discrete bulk spectrum with random statistics appears naturally in the Lorentzian 
description of Jackiw-Teitelboim (JT) gravity if an extra confining potential is introduced in the
region where the renormalized geodesic length becomes of order $e^{S_0}$.
The existence of such an extra confining potential may be inferred from the late behavior
of complexity and also from the Saad-Shenker-Stanford (SSS) duality between  JT gravity and
the matrix model. We derive the explicit form of the  extra confining potential from
the well-established density of states obtained in the Euclidean approach 
to  JT gravity. This extra potential is implicitly determined by the solution
of the Abel's integral equation which turns out to be identical to the 
string equation of the matrix model in the SSS duality formulation of 
JT gravity. Thanks to the  extra confining potential and the random 
nature of the spectrum, the time evolution of the Krylov complexity, which 
is identified with the renormalized geodesic length, naturally exhibits four 
phases, namely a ramp, a peak, a slope, and a plateau.
}

\end{titlepage}


\section{Introduction}\label{sec0}
Black holes, once considered purely theoretical, have now been extensively confirmed through astrophysical observations. Within the framework of general relativity, they are described by classical solutions and were not originally associated with the entropy of any physical system. However, when thermodynamic considerations are introduced, black holes are argued to possess a finite entropy, known as the Bekenstein-Hawking entropy \cite{Bekenstein:1972tm,Hawking:1974sw}.
Understanding its microscopic origin is an important challenge, as it is expected to shed light on quantum gravity beyond classical general relativity. In recent years, a two-dimensional model of dilaton gravity, known as Jackiw-Teitelboim (JT) gravity~\cite{Jackiw:1984je,Teitelboim:1983ux}, has been intensively studied, partly because it serves as a natural playground for exploring the AdS/CFT correspondence and allows for analytic treatments in a controlled setup to test ideas about quantum gravity, particularly in the context of black hole interiors.

Given that our spacetime has a Lorentzian signature, it is natural to adopt a Lorentzian framework for understanding quantum gravity. In particular, the Lorentzian signature is well-suited for black hole interiors, while there is no a priori  justification for the Euclidean approach in this context. Nevertheless, in the semiclassical limit, the initial state of quantum fields in a black hole geometry is often chosen through a Euclidean continuation, leading to the Hartle-Hawking state~\cite{Gibbons:1976ue}. This state represents a thermodynamic equilibrium of matters with a black hole and is regarded as a global state constructed on the two-sided geometry of black holes, which is identified with a thermofield double state (TFD) in the context of the AdS/CFT correspondence~\cite{Maldacena:2001kr}. In JT gravity, one may go beyond the semiclassical approach and incorporate quantum effects from gravity. Since local bulk gravity has no local degrees of freedom, this simplifies the analysis greatly.

Intriguing results obtained through the Euclidean path integral method in JT gravity include the disk partition function~\cite{Stanford:2017thb}, the correspondence with random matrix integrals~\cite{Saad:2019lba},
the replica wormhole computation for generalized entanglement entropy~\cite{Penington:2019kki,Almheiri:2019qdq}, and spectral form factors~\cite{Cotler:2016fpe},  among others. For a recent review, see~\cite{Turiaci:2024cad,Mertens:2022irh}. In particular, Saad, Shenker and Stanford~\cite{Saad:2019lba}  established the correspondence between JT gravity and a random matrix model, referred to as the SSS duality in the following. This duality suggests that the boundary theory dual to JT gravity may consist of random ensembles rather than a single theory. 
While Euclidean results have provided valuable insights into quantum JT gravity, it would be much more satisfactory if a Lorentzian approach could reproduce the Euclidean results.

The Lorentzian picture also shows 
the growth of the black hole interior or the wormhole length (volume) under forward global time evolution. This growth phenomenon is conjectured to provide the bulk realization of the increasing computational complexity of the boundary theory~\cite{Susskind:2014rva, Susskind:2014moa,  Susskind:2018pmk}.  

In JT gravity, the correspondence between geometric quantities and complexity has been made more concrete. For example, in~\cite{Iliesiu:2021ari}, the renormalized bulk geodesic length between two boundary points at late times is argued to match the complexity of the boundary theory. However, this match requires a specific prescription to renormalize the geodesic length through the wormhole, which remains unclear from the perspective of  Lorentzian boundary Schwarzian theories.  Another question from the Schwarzian perspective concerns understanding the density of states computed from the disk partition function in Euclidean JT gravity or from the Euclidean Schwarzian theory, given that the Schwarzian theory corresponds to a quantum system with a continuous spectrum~\cite{Stanford:2017thb}.

To reconcile the Euclidean and Lorentzian approaches in JT gravity, we propose a simple fix in this paper: introducing an extra confining potential 
to  the quantum mechanical system of 
the 
two-sided Schwarzian theory, which becomes relevant only
in the
region where the renormalized geodesic length becomes of order $e^{S_0}$. Essentially, this extra confining potential is designed to match the density of states obtained in the Euclidean approach,
which obviously makes the bulk spectrum discrete. 
Additionally, it ensures the complexity to evolve as expected.
Specifically, 
at early times, the renormalized length 
grows linearly and reaches a maximum due to our extra confining potential. Then, it starts to decrease and eventually stabilizes at a plateau value at late times. This behavior is quite consistent with numerical results for the Krylov complexity of the boundary system~\cite{Balasubramanian:2022tpr,Balasubramanian:2019wgd}.  
Furthermore, we show that our extra confining potential is identical to the one 
appearing in the so-called string equation for JT gravity~\cite{Johnson:2019eik}. 
However, the contexts in two 
approaches that lead to the same expression turn out to be 
quite different. 

This paper is organized as follows. In Section~\ref{sec1} we review  the quantum mechanical system derived from  the two-sided Schwarzian in JT gravity and the SSS duality establishing our conventions. Section~\ref{sec2} outlines our reasoning behind  introducing an extra confining potential for the quantum mechanical system. We present some computational details to determine our extra confining potential in Section~\ref{sec3}. In Section~\ref{sec4} we review the string equation from the matrix model dual to JT gravity, confirming that an identical potential with ours  also arises in this context. 
We also comment on the incompleteness of the Hamiltonian or potential in the string equation, highlighting the main differences between our quantum mechanical system and the auxiliary system in the string equation. In Section~\ref{sec5}, we present arguments for the origin of the plateau behavior of the expectation value of the position operator in a TFD state. This is a direct consequence of the cancellation of the rapidly oscillating phase of the TFD state at late times. We also provide numerical results on the expectation value of the position operator in this section. Finally, we summarize our results and outline future directions in Section~\ref{secCon}. In Appendix~\ref{AppA}, we provide an alternative derivation of  our extra confining potential directly from the disk partition function. 


\section{JT gravity and SSS duality}\label{sec1}
In this section we review the boundary Schwarzian aspect of JT gravity~\cite{Almheiri:2014cka,Kitaev2014,Maldacena:2016hyu,Maldacena:2016upp,Jensen:2016pah,Engelsoy:2016xyb,Mertens:2017mtv,Kitaev:2017awl} in association with the SSS duality, which serves as our setup in the following quantum mechanical interpretation  of bulk physics.   For JT gravity and its boundary Schwarzian formulation,  we adopt the two-sided Lorentzian picture  while employing frequently the Euclidean picture  to incorporate the SSS duality. Let us begin  with the JT gravity action, given by:
\begin{equation} \label{}
I = \frac{1}{16\pi G}\int_{M} d^{2}x\sqrt{-g}~\phi (R+2) + \frac{1}{8\pi G}\int_{\partial M}du \sqrt{-\gamma_{uu}}~  \phi (K-1) \,,
\end{equation}
where $\phi$ is a dilaton field, $u$ represents the boundary time,  and $\gamma_{uu}$ and $K$ are the induced metric and the extrinsic curvature on the boundary $\partial M$, respectively.   Here, the surface term consists of the left and right parts in the  two-sided  Lorentzian framework. It is well-known that these boundary actions can  be interpreted as describing the fluctuating  boundary dynamics of  the left and right boundary particles.  In the following, we set $8\pi G=1$.

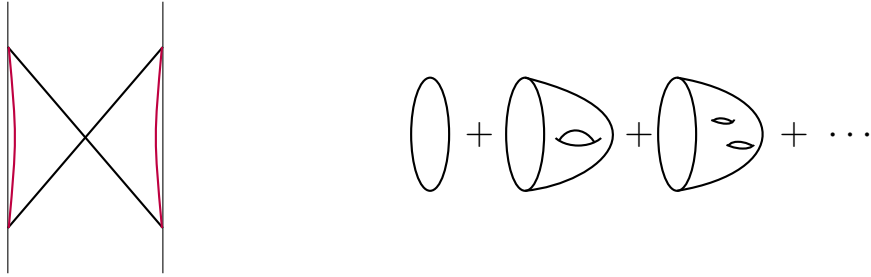
\begin{figure}  
\vskip0.3cm 
\begin{center}
\begin{multicols}{2}
%
\begin{tikzpicture}[scale=1.2]
\draw
(1.7,0)--++(0,3);
\draw
(3.4,0)--++(0,3);
\draw[thick] (3.4,0.5) -- (1.7,2.5) ;
\draw[thick] (1.7,0.5) -- (3.4,2.5) ;

\draw[thick,purple] 
(1.71,0.5) .. controls (1.8,1.5) .. (1.71,2.5);
\draw[thick,purple] 
(3.39,0.5) .. controls (3.3,1.5) .. (3.39,2.5);
\end{tikzpicture}

\vspace{1.2cm}

\columnbreak


\begin{tikzpicture}[scale=0.5]
 
\draw[white, thick,opacity=0] (-6,0) ellipse (0.5 and 1.5);
\draw[black, thick] (-6,0-2) ellipse (0.5 and 1.5); 

\node at (-4.7,0-2) {\large $+$};

\draw[black, thick] (-3.5,0-2) ellipse (0.5 and 1.5); 

\draw[black, thick] (-3.5,1.5-2) .. controls (0.5,0.5-2) and (-1.5,-1.3-2) ..   (-3.5,-1.5-2) ; 

\draw[black, thick]  plot [smooth, tension=1.2] coordinates {(-2.7,-0.1-2)  (-2.1,-0.3-2) (-1.5,-0.1-2) }   ;  
\draw[black, thick]  plot [smooth, tension=1.2] coordinates {(-2.6,-0.16-2)  (-2.15,0.1-2) (-1.68,-0.21-2) }   ; %

\node at (-0.5,0-2) {\large $+$};

\draw[black, thick] (0.5,0-2) ellipse (0.5 and 1.5); 

\draw[black, thick] (0.5,1.5-2) .. controls (3.5,1-2) and (3.5,-1.0-2) ..   (0.5,-1.5-2) ; 

\draw[black, thick]  plot [smooth, tension=1.2] coordinates {(1.45,0.36-2)  (1.68,0.42-2) (1.96,0.32-2) }   ;  
\draw[black, thick]  plot [smooth, tension=1.2] coordinates {(1.4,0.38-2)  (1.8,0.28-2) (2.0,0.39-2) }   ; %

\draw[black, thick]  plot [smooth, tension=1.2] coordinates {(1.8,-0.28-2)  (2.23,-0.38-2) (2.52,-0.28-2) }   ;  
\draw[black, thick]  plot [smooth, tension=1.2] coordinates {(1.85,-0.3-2)  (2.1,-0.2-2) (2.47,-0.3-2) }   ; %

\node at (4.5,0-2) {\large $+ ~\cdots$} ;
\node[opacity=0] at (4.5+3,0-2)  {\large $\color{white}+ ~\cdots$} ;













\end{tikzpicture}

\end{multicols}

\end{center} 
\vskip-0.3cm 
\caption{\small On the left we draw the left and right cutoff trajectories 
as purple curves near the AdS$_2$ boundaries, respectively.  On the right we illustrate the genus expansion with one boundary. }\label{fig1}
\end{figure}

The bulk equation of motion $R+2=0$ yields AdS$_{2}$ as the bulk solution, whose metric  in global coordinates is written as 
\begin{equation} \label{}
ds^{2} = \frac{-d\tau^{2}+d\mu^{2}}{\cos^{2}\mu}\,.
\end{equation}
By imposing the cutoff conditions on each boundaries with a cutoff parameter $\epsilon$ 
as
\begin{equation} \label{}
ds^{2}\Big|_{\partial M} = - \frac{du^{2}}{\epsilon^{2}}\,, \qquad \phi\Big|_{\partial M}  = \frac{\bar{\phi}}{\epsilon}\,,  
\end{equation}
the action  reduces to 
\begin{equation} \label{}
I =  \int du  \Big(L_{\ell} +L_{r}  \Big)\,, \qquad L_{\ell,\, r} = \frac{{\cal C}}{2} \bigg[ \Big(\frac{\tau''_{\ell,\, r}}{\tau'_{\ell,\, r}}\Big)^{2} - \tau'^{2}_{\ell,\,r}\bigg]\,,
\end{equation}
where the total derivative terms are discarded. Here,  ${\cal C}$ can be identified with $\bar{\phi}$,  which is also set to  $2{\cal C}=1$  in the following. (See the left figure of Figure \ref{fig1}.)
Following  standard procedures for higher derivative theories, one may introduce additional phase space variables $\chi_{\ell,\, r}$  as $\tau'_{\ell,\,r} = {\cal C}e^{\chi_{\ell,\, r}}$, for instance, through a Lagrange multiplier.  Then, one can obtain the Hamiltonian from the Schwarzian. However, the SL$(2,{\bf R})$ isometry of the AdS$_{2}$ background should be imposed as  gauge constraints in the Schwarzian formulation~\cite{Marolf:2000iq,Marolf:2008it,Harlow:2018tqv,Lin:2019qwu,Penington:2023dql}. For pure JT gravity, it turns out that the left and right boundary dynamics cannot be independent, since $H_{r} - H_{\ell}$ constitutes  one of the gauge constraints. On the other hand, the total  Hamiltonian $H_{tot} \equiv H_{\ell}  + H_{r}=2H$  leads to a  meaningful  dynamics, where the Hamiltonian $H$ is given by
\begin{equation} \label{totalH}
H_{l}  = H_{r}=  
H= p^{2} + e^{q}\,,
\end{equation}
where  $q \equiv \chi_{\ell} +\chi_{r}$ and $p$ is the canonical conjugate variable to $q$.

To incorporate the gauge constraints on a Hilbert space~\cite{Penington:2023dql,Penington:2024sum}, we should be careful due to the  non-compact nature of SL$(2,{\bf R})$, which may be handled  via the so-called  Hilbert space of coinvariants rather than invariants. On the physical Hilbert space ${\cal H} \simeq L^{2}({\bf R})$, $p$ acts as $p=-i\frac{d}{dq}$  satisfying $[p,q]=-i$. Notably, the renormalized geodesic length $\ell_{\gamma}$ between the left and right boundary trajectories specified by $\tau_{\ell}(u)$ and $\tau_{r}(u)$ is related to the $q$ variable as~\cite{Harlow:2018tqv,Lin:2022rbf,Bak:2023zkk}
\begin{equation} \label{}
\ell_{\gamma} = \ell_{bare} -\ln 2\phi|_{\ell} - \ln 2\phi|_{r} = -q\,.
\end{equation}
Under this identification,  the physical Hilbert space ${\cal H} \simeq L^{2}({\bf R})$ of pure JT gravity is given by the wave functions of the renormalized length $\ell_{\gamma}$.   As is clear from the Liouville type potential of  the total Hamiltonian in~\eqref{totalH}, the scattering state wavefunction is given by the modified Bessel function $K_{i 2\sqrt{E}}(2e^{-\ell_{\gamma}/2})$ and the spectrum should be continuous. As a result,  the density of states cannot be defined at this stage.

Since JT gravity can be regarded as a  certain low energy limit of a UV complete theory or a higher dimensional theory, it would be natural to consider a UV completion or a non-perturbative completion of JT gravity. 
In this context,  a complete non-perturbative theory 
would lead to a discrete spectrum with random statistics whose level spacing is
characterized by the scale $O(e^{-S_0})$. The  continuum limit  of such theory becomes pure JT gravity. 
One of such completion or extension is the identification of pure JT gravity with a ``double scaling'' limit of a specific matrix model, known as the SSS duality.  According to this duality,  the Euclidean JT gravity partition function on various topologies can be computed via the corresponding matrix integral formula.  Along this line, we may include the topological term  in the Euclidean action   $I_{E}$, corresponding to the finite entropy $S_{0}$,  as 
\begin{equation} \label{}
I_{E}^{top} \equiv  -\frac{S_{0}}{2\pi} \bigg[ \frac{1}{2}\int_{M_{E}} \sqrt{g} R  + \int_{\partial M_{E}}  \sqrt{\gamma} K \bigg]\,,
\end{equation}
which is used for the  topological expansion of the partition function in powers of $e^{-S_{0}}$. 

With  the Euclidean  topological term of   pure JT gravity,   the disk partition function for the boundary wiggles, with the renormalized circumference  $\beta$,  can be computed  for instance, by using the one-loop exact boundary Schwarzian action~\cite{Stanford:2017thb,Saad:2019lba}, leading to
\begin{equation} \label{}
Z_{disk} (\beta) = \frac{e^{S_0}}{4\sqrt{\pi}} \frac{1}{\beta^{3/2}} e^{\frac{\pi^2}{\beta}}\,.
\end{equation}
This partition function can be thought  to be related to the density of states $\rho_{JT}(E)$  which is given by
\begin{equation} \label{}
\rho_{JT}(E) = \frac{e^{S_{0}}}{4\pi^{2}}\sinh(2\pi\sqrt{E})\,.
\end{equation}
Since $\rho_{JT}(E)$ is a continuous function of $E$, it cannot be interpreted as a conventional density of states in a quantum mechanical system. 
However, in the following sections, we suggest a possible interpretation of $\rho_{JT}(E)$ as a density of states of a specific quantum mechanical system by introducing a confining potential, taking into account the $e^{S_{0}}$ dependence of $\rho_{JT}(E)$ explicitly. In this sense, $\rho_{JT}(E)$  may be denoted as $\rho_{JT}(E, e^{S_{0}})$.

To fix our notation and distinguish our main suggestion in later sections from the SSS duality, we shall give a very short review  of the double scaling limit of the matrix model, which is known to be ``solvable'' in the $L\to \infty$ limit.  See for a review of matrix models~\cite{Ginsparg:1993is,DiFrancesco:1993cyw,Guhr:1997ve}. The matrix model partition function for a Hermitian $L\times L$ matrix $H$, ${\cal Z}$ is given by
\begin{equation} \label{MatrixPart}
{\cal Z} = \int d H~ e^{-L\text{Tr}\, U(H)}\,.
\end{equation}
In the matrix model,  it is convenient to introduce the so-called resolvent $R(E)=\text{Tr}\, \frac{1}{E-H}$, which is related to the density of eigenvalues as $R(E-i\epsilon) - R(E+i\epsilon) = 2\pi i\rho(E)$.  Here, the density of eigenvalues is defined by $\rho(E) \equiv \text{Tr}\, \delta (E-H)= \sum\limits^{L}_{j=1}\delta(E-\lambda_{j})$, where $\lambda_{i}$ are  the eigenvalues of the Hermitian matrix $H$. 
Taking into account the Vandermonde factor in the Hermitian matrix model, the large $L$ saddle point  equation to the matrix model partition function in~\eqref{MatrixPart} is given by
\begin{equation} \label{Uprime}
U'(E)= \frac{2}{L}\dashint  d\lambda~  \frac{ \rho(\lambda)}{(E-\lambda)}\,,
\end{equation}
where the integral is over the principal value of the integrand.  For a given potential $U(E)$,  the equation~\eqref{Uprime} can be used to obtain the relation $R_{0,\,1}(E-i\epsilon)+ R_{0,\,1}(E+i\epsilon) = U'(E)$ for the zero genus  and one boundary resolvent $R_{0,1}(E)$.  To obtain $R_{0,1}(E)$, we need to choose branch cuts associated with the eigenvalue distribution of the potential $U(H)$  in~\eqref{MatrixPart}.  SSS proposed that a one-cut solution suffices to describe JT gravity, which may be chosen as $[0,2a]$ on the real line.   Introducing a function $\sigma(x) = x(x-2a)$, the end point $a$ can be determined from the potential $U(E)$. And then, $U'(E)$ and $R_{0,1}(E)$ give us the so-called spectral curve of the matrix model, which plays the role of initial data for the topological recursion relation of the resolvent correlators~\cite{Eynard:2004mh,Mirzakhani:2006fta,Eynard:2007kz,Eynard:2007fi}.

To relate JT gravity correlators to those of a  matrix model, SSS proposed  that the density of states for JT gravity can be understood as the ``double scaling'' limit of the  density of eigenvalues of a certain matrix model~\cite{Brezin:1990rb,Gross:1989vs,Douglas:1989ve}. This limit is defined by $L\to\infty$ and $a\to\infty$,  while keeping $e^{S_0}$ finite. As a specific density of eigenvalues, SSS has suggested the following expression: 
\begin{equation} \label{defa}
\rho(E,a)= \frac{e^{S_0}}{4\pi^2} \sinh 2\pi \sqrt{E\big(1-
E/2a
\big)} \,, \qquad L=\int^{2a}_0 \rho(E,a)\,,
\end{equation}
which  can be used to determine   $U(E,a)$  from the density of state $\rho(E,a)$ by the relation 
\begin{equation} \label{}
U'(E,a)= \frac{1}{L}\dashint^{2a}_{0} dE'\frac{ \rho(E',a)}{(E-E')}\,.
\end{equation}
 In the double scaling limit, one may identify the density of eigenvalues in the double scaled matrix model as $\lim\limits_{a\to\infty}\rho (E,a) = \rho_{JT}(E)$. Equivalently, the disk partition function in the matrix model is matched with that  in Euclidean  pure JT gravity. 
This identification allows us to use the matrix model techniques to obtain JT gravity correlators. As mentioned earlier, we explicitly  keep $L= e^{S_{0}} g(a)$ (with $g(a)$ specified by (\ref{defa})) to clarify our prescription for a confining potential introduced later. 
The integral of the leading density of eigenvalues, $\rho(E,a)|_{a\to\infty}$, without rescaling by $e^{S_{0}}$, yields a result of  order $e^{S_{0}}$. 

Basically, the SSS duality asserts that whole correlators of Euclidean  pure JT gravity partition functions are determined by the corresponding double scaled matrix model. 
Schematically, the primary focus in the topological expansion of the partition function with $n$ boundaries in the double scaled matrix model lies in the connected part of the resolvents 
\begin{equation} \label{}
\langle R(E_{1})\cdots R(E_{n})\rangle_{conn} \simeq \sum^{\infty}_{g=0} e^{-S_{0}(2g+n-2)} R_{g,n}(E_{1},\cdots, E_{n}) \,,
\end{equation}
which may be rewritten in terms of the correlators of the Euclidean JT gravity partition functions, $\langle Z(\beta_{1}),\cdots, Z(\beta_{n})\rangle_{conn}$. 
These correlators are known to satisfy  specific topological recursion relations and then related to Weil-Petersson volume of the moduli space of a genus $g$ surface with $n$ geodesic boundaries of length $b_{1}, \cdots, b_{n}$.  According to the SSS duality,  all such  correlators can be determined completely by two initial inputs:  disk partition function $Z_{disk}(\beta)$ and trumpet partition function $Z_{trumpet}(\beta, b)$. Both of these quantities are computed from Schwarzian boundary wiggles in Euclidean pure JT gravity.

While the SSS duality provides a UV completion of Euclidean pure JT gravity, its matrix model formulation complicates the direct visualization of the discrete spectrum with random statistics.   In the following sections, we present a simple quantum mechanical interpretation of the  density of  states $\rho_{JT}(E)$, starting from the boundary 
Schwarzian Hamiltonian in~\eqref{totalH}. By introducing a confining potential in Lorentzian picture, we interpret the density of state $\rho_{JT}(E)$ in Euclidean pure JT gravity as a leading continuum limit obtained when the potential  vanishes. Using this framework, we also explore a consistent interpretation of  quantum state  complexity of time evolution in terms of the bulk renormalized geodesic length.

\section{Confining Potential}\label{sec2}
The needs for a left  confining potential   ({\it i.e.}    for $-q \gg 1$) may be argued in the following ways. First of all,
this potential naturally follows from the late behavior of complexity shown in~\cite{Balasubramanian:2019wgd,Iliesiu:2021ari,Balasubramanian:2022tpr}. 
Note that the complexity operator may be identified with $ \ell_\gamma=-q$ 
where $\ell_\gamma$ is the geodesic length~\cite{Susskind:2014rva,Brown:2015lvg,Brown:2015bva,Miyaji:2015woj,Mertens:2017mtv,Brown:2018bms,Susskind:2020gnl,Iliesiu:2021ari,Lin:2023trc,Iliesiu:2024cnh,Boruch:2024kvv,Miyaji:2024ity}. With the  above  Hamiltonian {\color{blue}   in~\eqref{totalH}}, we have
\be
 \frac{1}{2}\frac{d^2}{d t^2} \langle q \rangle_{tfd} = - \langle e^q \rangle_{tfd}\,,
\ee
where the TFD state will be specified below explicitly. Initially, there is a negative effective 
force since the force term in the right hand side         ({\it i.e.} $q \gtrsim 0)$ is negative definite. 
As $-q$ becomes large ({\it i.e.} $-q\gg 1$), the force in the right side becomes negligible and 
\be
- \langle q \rangle_{tfd} \sim C_1\, t
\ee
with $C_1$ to be an O(1) positive coefficient~\cite{Yang:2018gdb}. 
Even including  the perturbative and nonperturbative contributions\footnote{Here perturbative  
contributions are given in terms of genus expansion counted by even powers of
$e^{-S_0}$ which corresponds to $e^{-\# 1/G_N }$ in the gravity side. Then the nonperturbative contribution are order of  $O(e^{-\# e^{S_0}})$ which is extremely small if the   number $\#$ is positive. However, even if taking $e^{S_0}$ to be very large, these nonperturbative contributions may not be ignored when the number $\#$ becomes imaginary.}, the above behaviors continue until $t\ll e^{S_0}$. 
In Ref. \cite{Iliesiu:2021ari}, it was further shown that
\be
- \langle q \rangle_{tfd} \rightarrow e^{S_0} C_2  \qquad  \text{as} \quad t \gg e^{S_0}\,,
\ee
where $C_2$ is another  O(1) positive coefficient, which has a nonperturbative nature.
We view that this behavior is arising from some complicated left confining potential that is relevant
as $-q$ becomes $O(e^{S_0})$.  Interestingly, this confining behavior 
 is 
originated 
from the purely Lorentzian 
 regime as the geodesic length of the wormhole becomes large in the large $t$ region. 

From the view point of the Lorentzian   boundary quantum mechanics, any higher genus contributions, which
are either perturbative or nonperturbative,
mainly happen deep inside the bulk and hence affect the large $\ell_\gamma\, (=-q)$  behaviors,
which may be summarized in terms of the left confining potential. 

Another argument is from the SSS duality between JT gravity and the matrix model
where the matrix  has a size $L = e^{S_0}g(a)$. The density of states is defined by
\be
\rho(E) = \sum^L_{n=1}  \delta (E-E_n)\,,
\ee
and it is  a fundamental quantity from which the partition function follows automatically by
the relation
\be 
Z(\beta)= \int dE \rho(E)  e^{-\beta E}\,.
\ee 
Below, we will be mainly interested in the corrections  of quantum mechanics
which becomes $O(1)$ when $-q$ becomes of order $e^{S_0}$ and are not interested in 
the simple perturbative corrections of $O(e^{-S_0})$ which changes $\rho_0(E)$ by 
$O(e^{-S_0})$.    
Such corrections can be summarized by an effective left confining potential in the boundary quantum mechanics.  The existence of this potential stems from the defining properties of the
matrix model in the matrix model side.

\begin{figure}[htbp] 
\begin{center}
\begin{tikzpicture}[scale=1.8]
\begin{axis}[
        axis lines =  center ,
        axis y line=none,
        axis x line=none,
        xlabel = $x$,
        ylabel = {$V(q)$},
        xmin=-20, xmax=5,
        ymin=-0.3, ymax=1,
        xtick=\empty,
        ytick=\empty,
        xlabel style={below right},
        ylabel style={left above},
        clip=false,
        y=0.8cm
     ]
       
    \addplot[blue, smooth, domain=-15:4.6, samples=200] {250*exp(x-9.8)+0.015};
     \node[font=\tiny] at (axis cs:5.5,1.4)  {$e^{q}$};
     \addplot[blue, smooth, domain=-20:-14.98, samples=200] {-2.4*log10(log10(x+22.5))- 0.13};
    \draw[->] (axis cs:2.5,-0.2) -- (axis cs:2.5,1.6)  ;
    \node[font=\tiny] at (axis cs:2.1,1.68)  {$V$};
     \draw[->] (axis cs:-20,0) -- (axis cs:5,0)  ;
     \node[font=\tiny] at (axis cs:5.3,-0.15)  {$q$};


    \draw[dotted,thick] (axis cs:-18.5,0) -- (axis cs:-18.5,0.4);
   
    \node[font=\tiny,below] at (axis cs:-17,0) {$q= {\cal O}(-e^{\text{$S_{0}$}}) $};
    
 %
\end{axis}
\end{tikzpicture}

\end{center}
\caption {A schematic form of the potential $V(q)=e^q+W(q)$, where the left confining potential $W(q)$  becomes $O(1)$ only when  $q$ becomes of $O(-e^{S_0})$. With the left confining potential, the spectrum becomes discrete. 
}\label{fig2}
\end{figure}
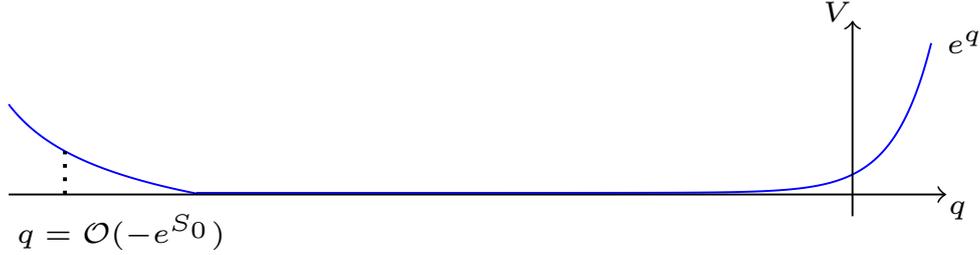

Interestingly with the left confining potential, by taking $e^{S_0}$ large but finite, the spectrum naturally becomes discrete. In the following, we assume the existence of the left confining potential   and determine its form explicitly.  Namely the total potential is given by
\be
V(q)=e^q + W(q)\,,
\ee
where the left confining potential $W(q)$ becomes $O(1)$ only when $q$ becomes of $O(-e^{S_0})$. As $e^{S_0}$ goes to infinity, the effect of the left confining potential disappears completely leading to the continuous spectrum we have in the beginning. A schematic form of the potential is illustrated in Figure 
\ref{fig2}.

Here we assume 
a Hamiltonian of the form $H=p^2+ V(q)$. Our approach in this note 
 may be
related to the formulation of matrix theory based on the so-called string equation. However, we do not aim to provide a precise formulation for deriving perturbative and nonperturbative corrections, as this is not our primary goal in this note.

\section{Discrete Density of States}\label{sec3}
In this section we would like to obtain the shape of the left confining potential $V(q)$ which reproduce
the desired density of states
\be 
\rho(E)= e^{S_0} \hat{\rho}(E)\,,
\ee
where $\hat{\rho}(E) =\frac{1}{4\pi^2} \sinh 2\pi \sqrt{E}$ to the leading order~\cite{Stanford:2017thb,Mertens:2017mtv,Bagrets:2016pxi,Saad:2019lba}.  As is well known, the density of states 
in the semiclassical limit with the left right confining potential is given by
\be \label{semidos}
\frac{1}{\pi}\frac{d}{dE} \int^{q_+}_{q_-} dq \sqrt{E-V(q)} = e^{S_0}\hat\rho (E)
\ee
where the left and right turning points $q_{\mp}$ are defined by the relation $E=V(q_\mp)$ with $q_+ >q_-$. 
Here the semiclassical limit is taken by sending $e^{S_0}$ to be very large. 
This relation directly follows from the so-called Bohr–Sommerfeld quantization condition. 
Let us 
briefly review the derivation of this result before proceeding further.
Using the semiclassical phase space approximation, we have
\be
\rho(E)= {\rm Tr} \, \delta (E-H) \simeq \frac{1}{2\pi} \int dq dp  \, \delta (E-H(p,q))\,.
\ee
We then make a canonical transformation to the
action-angle variables $(J,\theta)$ where $J$ is given by
\be
J(E)=\frac{1}{2\pi}\oint p dq=\frac{1}{\pi} \int^{q_+}_{q_-} dq \sqrt{E-V(q)}\,,
\ee
and the canonically-conjugated variable $\theta$ is ranged over $[0, 2\pi]$.  
The angle variable satisfies 
\be
\frac{d\theta}{dt} = \frac{dE(J)}{dJ}=\omega (J)\,,
\ee
where we used the relation $H(J)=E(J)$.
With these variables, the density of states becomes
\be
\rho(E)=\frac{1}{2\pi}\int d\theta dJ~ \delta (E-H(J))=\frac{dJ(E)}{dE} =
\frac{1}{\pi}\frac{d}{dE} \int^{q_+}_{q_-} dq \sqrt{E-V(q)} 
\ee 
in the semiclassical limit.  Thus we are led to the relation in (\ref{semidos}).

Now let us consider the potential of the form
\be \label{potentialV}
V(q)=e^{q} + W(X(q))
\ee
where we introduced
$X(q)=-e^{-S_0} q$ for later convenience.
We require $W(+\infty)=+\infty$ to ensure the left-confining property of the total potential.
On the right-hand-side, 
we do not want to ruin the original Liouville 
potential $e^q$; thus, we require, for the large positive $q$, $e^q \gg W(-e^{-S_0} q )$.
We shall also require $W(-\infty)=0$, so that the Liouville potential remains unchanged as $q$ becomes infinity\footnote{This potential also includes the random potential component that is responsible for a spectrum with random statistics and the corresponding ensemble average in the matrix model side. For instance, it may be organized as
\be\label{noise}
W=W_0(X) +\sum^\infty_{k=1} e^{-k S_0}v_{(k)}(q) W_{(k)}(X)\,,
\ee
where $v_{(k)}(q)$ denotes a Gaussian random noise function with $O(1)$ correlations, and $W_0(X)$  represents the saddle-point contribution,  which will be our primary concern in this note. By integrating out $v_{(k)}(q)$, one gets
\be
W_{\rm eff}(X)= W_0(X) +\sum^\infty_{g=1}e^{-2g S_0} \widetilde{W}_{(2g)}(X)\,.
\ee
These, however, are not derived from the first principle and therefore warrant careful examination with possible    modification required to ensure their consistency and validity.}.
We  now 
 assume that the function $W(X)$ does not explicitly depend on $e^{S_0}$,  restricting our attention to the  leading order contribution.
With this, 
one finds 
$q_+ = O(1)$ and $- q_-= O(e^{S_0})$. Then (\ref{semidos}) may be reduced to
\be 
\frac{1}{\pi}\frac{d}{dE} \int^{X_0}_{0} dX \sqrt{E-W(X)} = \hat\rho (E) \,,
\ee
where $W(X_0)=E$ and  we  ignored the $O(e^{-S_0})$ contribution as a leading approximation.  This then becomes
\be
\frac{1}{2\pi}\int^{E}_0 dY \frac{F(Y)}{\sqrt{E-Y}} =
 \hat\rho (E)\,,
\ee
where 
\be
F(Y)=\frac{dX}{dY}= \frac{1}{W'(W^{-1}(Y))} 
\label{fy}
\ee
with $W^{-1}(Y)$ denoting the inverse function of $W(X)=Y$.
This is nothing but the famous Abel's integral equation. The solution is given by
\be
F(Y)=\frac{d}{dY} \int^{{Y}}_{0} dE \frac{2 \hat\rho (E) }{\sqrt{Y-E}} 
\ee
Note the relation (\ref{fy}) implies that $dX = F(Y)dY$. Then this leads to
\be
X= \int^{{Y}}_{0} dE \frac{2 \hat\rho (E) }{\sqrt{Y-E}} =\frac{1}{\pi^2} \int^{\sqrt{Y}}_{0} ds \frac{s \sinh 2\pi s}{\sqrt{Y-s^2}}\,,
\ee
which can be integrated to give
\be \label{potential}
2\pi X=
\sqrt{W(X)} I_1 (2\pi \sqrt{W(X)})
\ee
with $Y$ replaced by $W(X)$.
Hence the potential $W(X)$ can be given implicitly with the above relation. In Appendix {\color{blue}A}, we will derive again this potential in an alternative way starting from the disk partition function as a validity check
of our method.  
In fact this equation is called the string equation~\cite{Okuyama:2019xbv,Johnson:2022wsr,Johnson:2020lns}
 in the matrix model formulation, which will be explained in the next section, (See~\cite{Balasubramanian:2022dnj} for a different  perspective on related matters.)

Using the 
fact that $I_1(x)= x/2 +{\cal O}(x^3)$ for small $x$, one finds
\be
W(X)= 2X + {\cal O}(X^2)
\ee
for small $X$. For large $X$, we use the asymptotic expansion
\be
 I_1(x) \simeq \frac{1}{\sqrt{ x}} e^{x} \Big(1+ {\cal O}(1/x)\Big)
\ee
for the large $x$.  This leads to
\be
W(X)=\left[  \frac{1} {2\pi } \ln \big((2\pi)^{3/2} X\big)  \right]^2 (1+ O(\ln\ln X/\ln X))\,.
\ee
Hence in the large $-q$ region, the slop $dW/dq$ of the potential behaves as
\be
\frac{dW(X(q))}{dq} \sim -\frac{e^{-S_0}} {\pi X} \ln \big((2\pi)^{3/2} X\big) \,,
\ee
which becomes extremely small.

The theory will be defined through the ensemble average 
\be \langle E_n \rangle= \int  {\cal D}\vec{v}(q) \, {\cal P}(\vec{v}(q)) E_n(\vec{v}(q))\,,
\ee 
where $E_n({\vec{v}(q)}) (n=1,2, \cdots)$ denotes  the $n$-th energy eigenvalue for a given collection of  random noise functions $\vec{v}(q)=(v_{(1)}(q),v_{(2)}(q) ,\cdots)$ in $W$ of (\ref{noise}), and the weight ${\cal P}(\vec{v}(q))$   should be determined by the original gravity theory. Of course, computing $ {\cal P}(\vec{v}(q))$ requires further study. However, for the discussions below, this ensemble average does not play an essential role, and we shall pick a particular sample when considering the simulation of the corresponding Schrödinger equation.

\section{String equation}\label{sec4}

In this section, we consider how the string equation originally appears in the double-scaled limit of the matrix model and compare it with our approach above. We begin with a brief review of the formulation \cite{Brezin:1990rb,Gross:1989vs, Douglas:1989ve, Okuyama:2019xbv,Mertens:2020hbs,Johnson:2020lns,Johnson:2020heh,Okuyama:2020qpm,Johnson:2020exp,Maxfield:2020ale,Okuyama:2021eju,Jafferis:2022wez,Johnson:2022wsr}. To evaluate the matrix integral, we use the technique of orthogonal polynomials, which satisfy
\be
\int^\infty_{-\infty} \, d\lambda\, e^{-L U(\lambda)} P_n (\lambda) P_m(\lambda)
=\delta_{mn} h_n \,,
\ee
where $P_n(\lambda)$ is an $n$-th order polynomial with $P_n(\lambda)= \lambda^n +\cdots$.
Assuming the potential $U$ is even, one finds $\lambda P_n= P_{n+1}+R_n P_{n-1}=B_{nm}P_m$, where 
the second equality defines the matrix $B$. The coefficient $R_n$ is the key quantity of interest in the matrix model and one may show that $R_n=h_n/h_{n-1}$ straightforwardly. Note that
\be 
\int^\infty_{-\infty} \, d\lambda\, e^{-L U(\lambda)} P'_n (\lambda) P_{n-1}(\lambda)
=L \int^\infty_{-\infty} \, d\lambda\, e^{-L U(\lambda)} U'(\lambda) P_n (\lambda) P_{n-1}(\lambda)=n\, h_{n-1}\,.
\ee
This leads to 
\be
Y_n =U'(B)_{n\, n-1} 
\ee
with $Y_n=n/L$. This equation in the continuum limit with $L\rightarrow \infty$ can be written as
\be 
Y= {\cal W}(R) +\frac{1}{6}\W^{[2]} (R) R\partial^2_Y R \, \epsilon^2+O(\epsilon^4)\,,
\label{ostring}
\ee
where $\W(R)$ can be found from the potential $U$ by the relation
$\W(R)=\oint  \frac{dz}{2\pi i} U'(z+R/z)$ and $\epsilon=1/L$ \cite{Ginsparg:1993is}. The higher order contributions in $\epsilon$ can be worked out explicitly order by order.  If $\W$ is given alternatively, again $U$ can be 
determined uniquely.  We introduce a double scaling limit by
\be
Y=1+(y-\mu) \delta^2, \ \ \ R=R_c -\sqrt{R_c}\, 
u(y) \delta^2
\ee 
with $L \rightarrow \infty$ and $\delta\rightarrow 0$ while keeping 
$\hbar=R_c^{1/4}
\epsilon /\delta^3=e^{-S_0}$ finite.  We require $\W(R_c
)=1
$. 
Further requiring
$t_n =(-)^{n-1} \W^{[n]} (R_c
) \delta^{2n-2} R_c^{n/2}
/n!$ for $n>0$ to be finite in the double scaling 
limit, one finds the string equation\footnote{Alternatively,  one may obtain the same string equation by taking  $\mu=0$ and $\W(R_c)=1-t_0 \delta^2
$ while keeping $t_0$ finite in the double scaling limit.}\negthinspace\negthinspace, 
\be
y+\sum^\infty_{k=0} t_k R_k (u)=0\,,
\label{streq}
\ee
where $t_0=-\mu$ with $R_0=1$. The function $R_k$ satisfies the recurrence relation  
\be
R'_{k+1}=\frac{2k+2}{2k+1}\Big( u 
R' _k+\frac{1}{2}u' R_k -\frac{\hbar^2}{4} 
R'''_k\Big)\,,
\label{Rrecc}
\ee
where the primes denote $y$-derivatives.
From (\ref{ostring}), one can check 
$R_0=1, \ R_1=u$ and $\  R_2 = u^2 -\frac{\hbar^2}{3\,\,} u''$ which are consistent with the above recurrence relation in (\ref{Rrecc}). As a solution of (\ref{streq}), the potential $u(y)$ can be expanded as $u(y)=\sum^\infty_{g=0} u_g(y)\hbar^{2g}$, which corresponds to the genus expansion of the matrix theory.
Note that in the $\hbar \rightarrow 0$ limit,
the string equation reduced to
\be
y+\sum^\infty_{k=0} t_k u_0^k=0 \,.
\ee
Let us introduce an orthonormal basis $\psi_n(\lambda)=\frac{1}{\sqrt{h_n}}e^{-L U(\lambda)/2} P_n(\lambda)$. Then the operator $\lambda$ is represented by
\be
\lambda \psi_n= \sqrt{R_{n+1}} \psi_{n+1} + \sqrt{R_{n}} \psi_{n-1} \,.
\ee
Hence one finds 
\be 
\lambda= \sqrt{R(Y+\epsilon)}e^{\epsilon \partial_Y} +\sqrt{R(Y)}e^{-\epsilon \partial_Y}.
\ee
In the double scaled limit, this leads to
\be
\lambda=2 \sqrt{R_c} 
-{\cal H}\,\delta^2 \,,
\ee 
where ${\cal H}= -\hbar^2 \partial_y^2+ u(y)$. The potential $u(y)$ can be obtained by solving
the string equation once $t_k$ is given. We introduce
the eigenstate 
\be
{\cal H}\,\psi(y,E)= E \,\psi(y,E) \,,
\ee
which is normalized by
\be
\int^\infty_{-\infty} dE \psi(y,E) \psi(y',E) =\delta(y-y') \,.
\label{psinor}
\ee
In fact, this wavefunction may be obtained with
\be
\psi_n\big(\lambda=2\sqrt{R_c}
-E \delta^2\big)L^{1/2}\delta^2  \rightarrow \psi(y,E)
\label{dswave}
\ee
by taking the double scaling limit with $\frac{n}{L}=1+(y-\mu) \,\delta^2$.
From this, the original may be computed as\footnote{Strictly speaking, one needs here an appropriate transformation to get back to the matrix model variables in Section \ref{sec1}. The relevant scaling symmetry reads $y\rightarrow s y$, $u \rightarrow u/s^2$, $t_k\rightarrow s^{2k+1}t_k$, ${\cal H}\rightarrow {\cal H}/s^2$, $\beta\rightarrow s^2 \beta$ and $\hbar\rightarrow \hbar$. Then one needs the scaling transformation with $s=\delta$ to get back to the variables in Section \ref{sec1}.} 
\be
\langle Z(\beta) \rangle =\int^\infty_{-\infty} dE \rho(E) e^{-\beta E}
=\int^{\mu}_{-\infty} dy\, \langle y| e^{-\beta {\cal H}}|y\rangle \,.
\label{MMdef}
\ee
From this, one further finds
\be \label{psiMM}
\rho(E)=\int^{\mu}_{-\infty} dy \, |\psi(y,E)|^2  \,.
\ee
Starting from this, one may determine any perturbative correlation functions to all orders in the double-scaling limit. Namely $\psi(y,E)$ contains all the information of the matrix model in the double-scaling limit.  

As mentioned in  \cite{Johnson:2022wsr}, this quantum mechanics problem is rather unconventional and unphysical.  To have the second equality in (\ref{MMdef}) ({\it i.e.} to have
the relation $\int^\mu_{-\infty}dy |y\rangle \langle y|=\int^\infty_{-\infty}dE |E\rangle \langle E|$), the eigenvalue problem for
${\cal H}\psi(y,E)=E \psi(y,E)$ in the conventional framework of quantum mechanics must be supplemented by a self-adjoint boundary condition at $y=\mu$, which is, in fact, NOT the case here. 
If a boundary condition or some wall potential were imposed, the spectrum would become discrete, as the configuration space would then be confined to the region $y \in (-\infty, \mu]$. However, it is clear that 
$E$ is continuous because it is defined by 
$\lambda=2
-E \delta^2$ and the variable $\lambda$ is continuous from the beginning.
The normalization of $\psi(y,E)$ defined by (\ref{psinor}) is also unconventional.
 The double-scaled wavefunction in (\ref{dswave}) never satisfies any particular boundary condition at $y=\mu$.
In fact,
\be
\rho(E) =\sum^{L-1}_{n=0} |\psi_n(\lambda=2\sqrt{R_c}
-E \delta^2)|^2L^2\delta^6 {\vert}_{\rm dslimit}=\int^{\mu}_{-\infty} dy \, |\psi(y,E)|^2 
\ee
and it is clear that $\psi_n(\lambda)L^{1/2}\delta^2$ does not satisfy any particular boundary condition. In the $L$ Fermion picture introduced in~\cite{Banks:1989df}, $y=\mu$ corresponds to the Fermi level. Indeed, in the evaluation of the two-point function of the operator, one needs information about the wavefunction in the region $y> \mu$, where particles appear instead of holes in the Fermi sea. Hence, the region above the Fermi level cannot be discarded, even in the double-scaling limit.

Now in the semiclassical limit with $\hbar \rightarrow 0$, one finds
\be
\rho(E)=\frac{1}{2\pi \hbar}\int^{\mu}_{-\infty} dy \frac{1}{\sqrt{E-u_0(y)}} \,.
\ee
With $\rho(E)=\rho_{JT}(E)$ and $\mu=0$, one finds
\be
y+\frac{\sqrt{u_0}}{2\pi}I_1(2\pi \sqrt{u_0})=0\,,
\ee
which agrees with our equation with the replacement $-y\rightarrow X$ and $u_0(y) \rightarrow W(X)$. This determines $t_k $ by
\be
t_k= \frac{\pi^{2(k-1)}}{2 k! (k-1)!} 
\ee 
with $k>0$ while $t_0=-\mu=0$~\cite{Okuyama:2019xbv,Mertens:2020hbs,Johnson:2020lns}.

As emphasized in \cite{Johnson:2022wsr}  and elaborated below  (\ref{MMdef}), the quantum mechanical framework here is merely a mathematical device organizing the matrix model computations and does not correspond to any physically sensible quantum mechanical system. To the best of our knowledge, no prescription is currently    available for rendering the corresponding system physical with a self-adjoint Hamiltonian. One might guess that the missing element is the Liouville potential $e^{y/\hbar}$, which would serve as the right-confining potential; however,  we do not see how this structure could arise naturally within the string–equation formulation.  By contrast, our confining bulk potential and the associated quantum-mechanical system are physical and, in this respect,  the two approaches are different. 
Nonetheless, the emergence of the same functional form for the left-confining potential within the string-equation formulation is quite striking, 
which suggests that the matrix model inherently encodes information about the dual bulk physics. 
One might expect that the left-confining potential arises precisely as a consequence of the matrix-model completion of JT gravity\footnote{We thank the referees for related remarks.}.


\section{Confining Potential and Complexity} \label{sec5}

 With the random potential, the spectrum follows the random statistics. The level spacing becomes order of $e^{S_0}$ showing the level repulsion. 
For a given Hamiltonian $H=p^2 + V(q)$ with $V(q)=W(X(q))+ e^q$, let us introduce  the eigenvalues $E_n$ with the corresponding eigenstate $|n\rangle$. Then the TFD state is given by
\be
|\psi(t)\rangle_{tfd} =\frac{1}{\sqrt{Z}}\sum_n e^{-(\frac{\beta}{2}+it) E_n}|n,n\rangle \,.
\ee
We have 
\bea
&& \frac{d}{d t} \langle q \rangle_{tfd} =-i \langle [q, H] \rangle_{tfd}= 2 \langle p \rangle_{tfd}\,,  \nn
&& \frac{d}{d t} \langle p \rangle_{tfd} =-i \langle [p, H] \rangle_{tfd}= - \langle V' \rangle_{tfd}\,.
\eea 
One  also has
\be
2 \langle p \rangle_{tfd}=-i \langle [q, H] \rangle_{tfd}
= \frac{i}{Z}\sum_{m,n} e^{-\frac{\beta}{2}(E_m+E_n)} e^{it(E_m-E_n)} (E_m-E_n)  
\langle m,m| q |n,n \rangle \,.
\ee
Then for the regime with $t \gg e^{S_0}$, the above becomes zero due to the 
random erratic phase cancellation, corresponding to 
the plateau region~\cite{Cotler:2016fpe,Susskind:2018pmk}: 
\be
\langle p \rangle_{tfd}\sim 0 \,,
\ee
and hence $\frac{d}{d t} \langle q \rangle_{tfd} \sim 0$ in this regime. This is also consistent
with the fact $\langle q \rangle_{tfd}$ becomes time independent in the regime
ignoring those erratic phase canceling contributions; this can be explicitly shown as
\be
\langle q \rangle_{tfd}=\frac{1}{Z}\sum_{m,n} e^{-\frac{\beta}{2}(E_m+E_n)} e^{it(E_m-E_n)} 
\langle m,m| q |n,n \rangle \sim \frac{1}{Z}\sum_{n} e^{-{\beta}E_n} 
\langle n,n| q |n,n \rangle
\ee
where the last relation follows by considering the phase cancellation of the middle expression
for $m\neq n$. In this regime, one may also show that  $-i \langle [p, H] \rangle_{tfd}= - \langle V'  \rangle_{tfd}\sim 0$ by the same argument.
As we have shown in the above, 
$F_-=-\langle e^q \rangle_{tfd} < 0$ becomes dominant in the beginning. As time goes,
the other (positive) contribution $F_+=-\langle W' \rangle_{tfd} > 0$ grows. As $F_-$ is getting negligibly smaller while $F_+$ still remains also negligible, $-\langle q \rangle_{tfd}$ grows linearly in time as was shown in the above.  As far as  $-\langle q \rangle_{tfd}$ grows, $F_-$ 
keeps decreasing while $F_+$ is increasing.
As $F_+$ becomes O(1), the growth of $-\langle q \rangle_{tfd}$ begins to decelerate.  It is rather clear that, at some point, 
$-\langle q \rangle_{tfd}$ reaches the maximum. At this maximum point,
it is clear that $F_+ > -F_-$ as $F_+$ contribution is winning over $-F_-$ for large $-\langle q \rangle_{tfd}$ as $-F_-$ becomes negligible. Then it turns around, $F_+$ is decreasing while $-F_-$ is growing. Eventually it stops as in the above where $\langle p \rangle_{tfd} \sim 0$ and $F_-+F_+ \sim 0$. (In principle, there could be some oscillations before reaching this point.) 

In the following, 
we perform a simple numerical study to illustrate the 
points discussed in the above. In particular, we numerically solve 
the Schr\"odinger equation,
\begin{equation} \label{schr}
i\partial_t \psi(q,t) = [-\partial_q^2 + V(q) ] \psi(q,t),
\end{equation}
and see the behavior of the expectation value of the complexity operator
$\langle \ell_\gamma(t) \rangle = - \langle q(t) \rangle$ as time evolves.
For the sake of illustration in this section, we incorporate the random part of potential as 
\be V(q)=e^q+W({\cal X}(q))\,,\ee 
with 
\be  \label{varvarX}
{\cal X}(q)=e^{-S_0} \big[\log (1+ e^{-q-\alpha}) +v(q)\big]\,,
\ee
where $\alpha$ is a random variable of order one, and $v(q)$ represents an $O(1)$ random potential.
To be specific,  the random number $\alpha$ is taken from the interval $[0,1]$ 
and we choose $v(q)$ of the form 
\begin{equation}
v(q) = v_0 \sum_{k=0}^{k_{max}} \frac1{k+1} \sin(p_k q + \alpha_k),
\end{equation}
where $v_0 \sim O(1)$, $k_{max} \gg 1$, $p_k$'s are some increasing sequence
of random numbers and $\alpha_k$'s are random phases. 

%
\begin{figure}[htb] 
\begin{center}
\includegraphics[height=5.5cm]{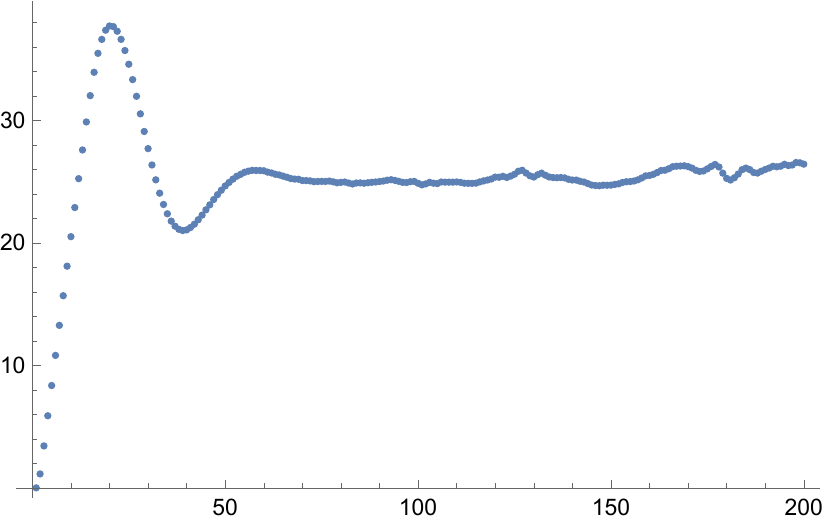}\phantom{aaa}
\end{center}
\caption{Complexity $\langle \ell_\gamma(t) \rangle = - \langle q(t) \rangle$ 
as a function of time. This figure is generated with the parameters
$S_0 = 4.5$,  $v_0=4.4$,  and $k_{max}=1000$ for definiteness, but the generic behavior is not 
changed with other values. The random number $\alpha$ is from the interval $[0,1]$.}\label{fig3}
\end{figure}
In Figure 
\ref{fig3}, 
we draw a typical time evolution of 
$- \langle q(t) \rangle$ for the initial wavefunction 
$\psi(q,0) = e^{-q^2/2}$ which is chosen to be a Gaussian wave packet
centered at the origin for simplicity.
As seen in the figure, the complexity $ -\langle q(t) \rangle $ initially 
increases linearly and reaches a peak which is due to the left confining potential 
$W$ in the far left region $-q \gg 1$. Then it turns around and start 
decreasing. It, however, does not return to the initial value but effectively 
stops at certain position at least up to the time of order $e^{S_0}$
because of the large dispersion of the initial wave packet due to the random 
potential $v(q)$. Thus, four phases of the complexity, namely a ramp, a peak,
a slope, and a plateau are naturally realized in this setup~\cite{Cotler:2016fpe,Saad:2018bqo,Balasubramanian:2019wgd,Balasubramanian:2022tpr,Erdmenger:2023wjg,Balasubramanian:2024ghv,Balasubramanian:2024lqk,Bak:2025qgs,Miyaji:2025yvm,Bhattacharyya:2025gvd}.

\section{Conclusions}\label{secCon}
In this paper, we have provided a mechanism by which the discrete bulk spectrum  with random statistics
is obtained naturally in the Lorentzian description of JT gravity.
In the quantum mechanical system derived from the two-sided Schwarzian
theory, we have introduced a left confining potential which is designed to
reproduce the density of states obtained in the Euclidean approach. 
It provides a very slowly growing wall which becomes important
in the region where the renormalized geodesic length becomes of order $e^{S_0}$.
The potential is implicitly determined by an equation which turns out
to be formally identical to the string equation of the random matrix model 
in the SSS duality formulation of JT gravity. However, the contexts in two 
approaches that lead to the same expression are apparently 
different. 
It would be an intriguing issue to further clarify 
any physical origin behind the coincidence.

By solving the Schr\"odinger equation of the quantum mechanical system,
we have obtained the time evolution of the renormalized geodesic length 
which corresponds to the Krylov complexity of the boundary theory.
We argued that the presence of the left confining potential, when combined with
the random nature of the spectrum, implies generically the emergence 
of a peak, a slope, and a plateau in the late time behavior
of complexity after the initial ramp. Since the concept of complexity
is not restricted to specific theories or spacetime dimensions, we expect
that the argument of this paper can be applied to more general cases.


In this paper, we have not attempted to derive the left confining potential by directly considering the perturbative or nonperturbative corrections to JT gravity. It would be interesting to determine whether this can be done in a physically motivated way. Here, we propose a very naive approach that leads to the full potential $W$, which is consistent with the full perturbative density of states. Let us introduce such a potential via  $ \int^\infty_{-\infty} dE \langle \rho(E)\rangle e^{-\beta E}=\int^\infty_{-\infty} dq \langle q| e^{-\beta H} |q\rangle $
where $\langle \rho(E)\rangle$ denotes 
the full perturbative density of states and $V(q)=e^q +W(-e^{-S_0} q)$ subject to the conditions
$\lim_{q \rightarrow \infty} W(-e^{-S_0} q) =0$ and $\lim_{q \rightarrow -\infty} W(-e^{-S_0} q) =\infty$. Of course, a random potential term should also be added, but we are uncertain whether this procedure is well-defined or not. Further studies are required in this direction.


\subsection*{Acknowledgement}
 D.B. was
 supported by the 2024 Research Fund of the University of Seoul. 
C.K. was supported by NRF Grant 2022R1F1A1074051.
 S.-H.Y. was supported in part by NRF grant funded by the Korea government(MSIT) RS-2025-00553127.

\appendix
{\center \section*{Appendix}}

\section{Alternative derivation of the potential $W$
} \label{AppA}
\renewcommand{\theequation}{\thesection.\arabic{equation}}
  \setcounter{equation}{0}
One can also derive the potential  $W$ from the disk partition function for JT gravity,
 which is given by
\be
Z_{disk}(\beta)=
\frac{e^{S_0}}{4\sqrt{\pi}} \frac{1}{\beta^{3/2}} e^{\frac{\pi^2}{\beta}}\,,
\ee
where we set ${\cal C}=1/2$.
 In the same approximation taken in (\ref{semidos}), this partition  function can be computed
 from the semiclassical phase space approximation  as
\be
Z(\beta)=\frac{1}{2\pi}\int dq dp \, e^{-\beta H(p,q)}\,.
\ee
 For the integral over the position variable $q$, we adopt the same assumption as in the
first derivation: We assume the form $V(q)=e^q + W(X)$  where $W(X)$ is an invertible function of $X(q)$ with $X(q) \in (0,\infty)$.  Under this assumption, the position integral becomes
\be
e^{-S_0}\int^{\infty}_{-\infty} dq\,  e^{-\beta V(q)} = \int^{\infty}_0 dX e^{-\beta W(X)} +O(e^{-S_0})\,,
\ee
where we ignored $O(e^{-S_0})$ corrections. 
We further note that
\be
 \int^{\infty}_0 dX e^{-\beta W(X)}= \beta \int^\infty_{0} d W X(W) e^{-\beta W}\,,
 \ee
 where we used the change of variables from $X$ to $W(X)$ and an integration by parts.
From this, one obtains 
\be 
\int^\infty_{0} d W X(W) e^{-\beta W}=\frac{1}{2\beta^2} e^{\frac{\pi^2}{\beta}}\,.
\ee 
By an inverse Laplace transform of the both sides, we  regain (\ref{potential}).
Note that in this alternative derivation, we used the disk partition function instead of
 the density of states applying different integral transforms. This appears to confirm the
 consistency of  our various assumptions.

\end{document}